\documentclass[journal]{IEEEtran}

\usepackage{algorithm}
\usepackage{algpseudocode}
\usepackage[export]{adjustbox}
\usepackage{graphicx}

\usepackage{subfigure}
\usepackage{tikz}

\usepackage{cite}

\begin{document}

\title{Transcoder Migration For Real Time\\Video Streaming Systems}

\author{Paul Farrow,
        Martin Reed,
        Maciej Glowiak,
        Joe Mambretti%
\thanks{P. Farrow is part of the Computer Science and
 Electrical Engineering Department, University of Essex, UK,
e-mail: pwfarr@essex.ac.uk.}%
\thanks{M. Reed is part of the CSEE dept. University of Essex, UK.}%
}

\maketitle

\begin{abstract}
The increase in real time ultra-high definition video services presents a challenging issue to current network infrastructures, because of its high bandwidth usage, which saturate network links. The required bandwidth is related to strict QoS requirements for digital media. There are systems in place currently to help reduce these problems, such as transcoders and application layer multicasting. However, these approaches are limited because they are usually implemented as  static resources. In contrast, by using the OpenFlow based system presented in this paper, it is possible to provide a more effective approach using dynamic resources - by both optimally placing transcoders in the network, as well as by migrating them to different locations while the streaming is taking place. This migration mechanism provides a near seamless switchover with minimal interruption to the clients.
\end{abstract}


\IEEEpeerreviewmaketitle

\section{Introduction}
\label{sec:introduction}
\IEEEPARstart{O}{penFlow} has allowed major advances in the way we can both manage and control traffic within a network. With its centralised controller architecture and its large range of capabilities, including packet header modification and packet redirection, it has the capability to change the way a `standard network' operates \cite{McKeown2008}. At the same time, advancements in multimedia technology have brought ultra-high definition content (\emph{e.g.} 4k) services to a larger audience \cite{mann2014ultra}, which has had a significant impact on the traffic utilisation of some network paths, especially those based on transoceanic cables. Bandwidth on these links will inherently become increasingly constrained as people begin to stream real time 4k content more frequently, which will require the use of systems that can minimise this added traffic. The required bandwidth is related to supporting strict QoS requirements for digital media, especially for high resolution content. This paper describes an approach to addressing this issue, based on using OpenFlow technology to control the delivery of ultra-high definition content.

Existing systems currently in operation attempt to minimise repeated traffic streams, while reducing the bandwidth that the streams consume. Two main mechanisms for achieving this are transcoders \cite{ahmad2005video} and application layer multicasting \cite{hosseini2007survey}. Transcoders can convert the media stream in various ways, such as by reducing bitrates, adjusting resolution, changing video format as well as many other functions. These techniques can be used to reduce the bandwidth consumed by the content stream, usually by sacrificing the visual quality of the content. However, this is not always the case as lossless compression techniques can also be utilised \cite{Andriani}. 

In the absence of good multicast support in IP networks, application layer multicasting can be used to reduce repeated content being sent through the network \cite{hosseini2007survey}. It achieves this reduction by distributing streams to multiple clients at positions closer to users than the location of the original sender. This permits a single stream to be sent through the network to the multicaster, which then sends the required multiple streams the remaining distance to the receiving clients. This type of stream branching can provide a significant reduction to the overall traffic load placed on the network while streaming 4K content, and it can work in parallel or individually with transcoders.

It is also important to mention the advancements in cloud technologies, as well as the proliferation of cloud resources around the world. Cloud based services have had a significant positive effect on software services, including the introduction to the concept of using virtual machines (VMs) as transcoding resources. This concept has many benefits over previous systems, such as CDN based systems which require the setup and use of dedicated machines and links placed throughout the network to perform transcoding services. A pool of dedicated transcoders may be an inefficient use of resources as it is unlikely that all transcoding machines would be fully utilised at all times. Another issue is that certain machines could be strained under heavy loads, for example, because they had not been configured since they were created. VMs however, which can be easily and quickly created, destroyed and migrated within multiple cloud datacentres \cite{Mishra2012}, have the ability to scale according to the load they are currently experiencing. This feature makes them a cost effective solution to providing transcoded content to geographically diverse clients. This functionality has facilitated scalability at a level that was previously not feasible for most applications. 

The proposition described in this paper is to use transcoders that are dynamically placed, according to need, using a cloud based computational resource. This basic proposition, while attractive, nonetheless has two inherent problems. However, both are addressed by the methods described in this paper. One is resolving the issue of where the transcoders should be best placed for efficient network utilisation. The other is resolving how traffic can be seamlessly switched between transcoders as the transcoder resources are moved. The transcoder placement problem needs to take into account that the general number of clients can move location over time, overloading networks in some areas and underutilising them in others. One solution to this would be to over-provision the network so that it is saturated with transcoders that could also provide application layer multicasting. This however is an inefficient use of resources, which would be costly to implement and maintain.
The solution outlined in this paper addresses both the transcoder placement optimisation as well as the mechanism for switching traffic between transcoders. The initial placement problem has been addressed by using a heuristic algorithm to optimise the placement of a given quantity of transcoders to reduce the total network load. This algorithm can be used in conjunction with a new mechanism developed to migrate transcoders in the network during transmission with minimal interruption to the client. This will enable the system to be optimised during a continued transmission with minimal disruption to viewers. The OpenFlow enabled switching mechanism developed for this is described and evaluated using experiments carried out using ultra-high definition video streamed over inter-continental connections.

The remainder of the paper is structured as follows: Section \ref{sec:background} provides relevant information on the topics of this paper. Section \ref{sec:transcoder_placement_optimisation} describes the transcoder optimisation problem, discusses the candidate solutions, provides information on the developed heuristic algorithm with the testing methodology and finally, presents the results obtained from the network model. Section \ref{sec:system_design} presents a detailed description of the OpenFlow enabled migration system. Section \ref{sec:results} then follows on from this to show results gathered from real world testing of both the OpenFlow enabled migration system alongside a system without the aid of OpenFlow for comparison. Finally, section \ref{sec:conclusions} presents the conclusions gathered from the results.

\section{Background}
\label{sec:background}
The solution presented in this paper requires a number of technologies and these are presented in this section. Although the solution presented may be used by any streaming system, it is particularly suited to future ultra-high definition systems as these require bandwidths far in excess of existing streaming systems. This section will present the components required to enable the underlying technology as well as presenting other solutions to the problem of streaming ultra-high definition video.

\subsection{Ultra-High Definition Video}
\label{sub:ultra_high_definition_video}
Video traffic has become one of the largest contributing factors to network traffic \cite{Colle}. There is increasing interest in the next generation of video systems, which incorporate technologies such as ultra-high definition video at resolutions of 4k or higher. Ultra-high definition video consumes a significant amount of bandwidth, typically in the order of 100 Mb/s for ultra-high quality compressed formats and up to 10 Gb/s in uncompressed form for 4k video \cite{Halak2010}. As with existing video, it is likely that next generation video will be consumed on heterogeneous playback systems and devices using varying access and core network services for delivery. Consequently, there is a need to modify a video source to suit the needs of the: end user, device, network and service provider.  Transcoding is the general term for systems that can modify coded media to suit alternative codecs and/or bitrate requirements \cite{Xin2005}. Generally, a transcoder accepts high-quality input and converts it to a lower-quality, reduced bit-rate output. One of the strong motivations for this work is focused on the issue of constrained network links where bandwidth availability is at a minimum or where it is available at high cost. Undersea links such as transatlantic cables are one example of such constrained links where bandwidth must be conserved when considering high-bandwidth applications such as next-generation video. 

\subsection{Transcoders}
\label{sub:transcoders}
Transcoding digital media on the Internet is an important step in the delivery of video content to end users; it provides not only a reduction in bitrate, so that more content can be delivered on a given set of links, but also allows the content to be tailored and adapted to the receiving device \cite{ahmad2005video}. This can include adapting the resolution of the video to match that of the client display and also transcoding to a format that the client can display.

The use of distributed transcoding is not a new area of study and early works in this area suggested including transcoding in the network layer forwarding components \cite{Sambe2005}. However, most of these early efforts violated the end-to-end principle that underpins the pragmatic deployment decisions of current IP based computer networks \cite{Saltzer1984}. However, with the more recent deployment of cloud systems and content delivery networks, end-system processing has become more of a consideration as part of a widely distributed architecture design that can draw upon other distributed resources to carry out storage or processing. Thus, we can enable distributed transcoding without violating the end-to-end principle of contemporary IP networks. Through the use of cloud architectures the transcoding becomes a flexible, relocatable, resource that is available to the application layer rather than an addition to network layer processing that is less flexible in nature.

Being able to move transcoding resources allows various delivery parameters to be optimised including latency and network bandwidth. This paper concentrates on a novel method of achieving network bandwidth resource savings. These savings can be achieved by reducing the number of streams between the transcoder and publisher. An efficient way of reducing repeated streams of traffic is to use the transcoder as a form of application layer multicasting. This technique allows the system to supply multiple resolutions of the same video to a large user base that are geographically widely dispersed and thus where IP multicast may not be available.

The transcoding resources can be considered as virtual machines which can be migrated live to servers in different geographical locations \cite{Yan}. Multiple transcoders can be moved at once with the implication of transferring a complete IP topology in a single movement operation.

This paper addresses both the optimisation of positioning the transcoding resources as well as building on this approach and detailing the mechanisms for dynamically moving transcoding resources.

\subsection{OpenFlow}
\label{sub:openflow}
OpenFlow is an SDN platform, which is becoming a popular tool for both research purposes as well as production networks \cite{McKeown2008}. It allows a large amount of switching resources to be controlled from a central controller, without sacrificing the scalability of per device control. This is achieved using a technique that allows flows to be set up in the switches by the controller as required, then the switches can function autonomously without controller intervention until a packet with no set rule is received; at this point the switch can then ask the controller for instructions on what to do. This separation of the control plane from the data plane (forwarding functions) allows great flexibility in the way a network can be operated, providing the ability to adjust the routing of packets dynamically as demands on the network change. 

Using this efficient process can provide a highly-scalable system, although with some challenges. For example a bottleneck may arise if a significant quantity of switches begin receiving unexpected traffic requiring intervention from the controller. This bottleneck can be minimised by setting up procedures for flows which would handle this unexpected traffic in other ways, such as dropping the traffic or redirecting to other switches in a hierarchical architecture.

Networking equipment manufacturers are starting to see the benefits of OpenFlow and are gradually implementing OpenFlow technologies in their network devices. One aspect that makes OpenFlow research and development so attractive, is due to how it is implemented in many networking devices such that it is possible to easily segment production traffic that is using standard network policies from OpenFlow enabled traffic flows \cite{McKeown2008}. This has allowed researchers to easily gain access to larger scale networks than would normally be available, since they do not require a dedicated OpenFlow research network. This segmentation capability has advanced the development of OpenFlow technology since it allows production traffic to travel through the network untouched by the OpenFlow rules, while still allowing researchers to have isolated OpenFlow enabled traffic redirection throughout the network for research and development purposes.

OpenFlow also has the ability to alter packet header fields during switching, which can be utilised to perform functions not possible with standard networking systems. Combining the dynamic flow insertion with the packet header manipulation makes OpenFlow a truly versatile networking system, which can be adapted based on the operators needs, without the restrictions of standard network operations. These are only some of the reasons that researchers and businesses are deploying OpenFlow enabled network equipment in their networks, creating the flexibility to implement and test ideas not previously possible.

\subsection{OpenFlow Controllers}
\label{sub:controllers}
OpenFlow enabled switches rely on controller applications to populate flow tables that are responsible for determining packet switching actions. This can be done both passively and dynamically, enabling switches to function without having fully populated flow tables at initialisation. Controllers are generally located at a centralised location to provide packet decisions to a large group of networking devices, this allows them to gain a good overview of the network topology enabling better packet routing decisions. The controller applications communicate with the OpenFlow switches with messages over a secure channel using the OpenFlow protocol, these messages include things such as packet received, packet actions and status messages.

Controllers, such as Floodlight \cite{floodlightWeb} provide a simple module based system where packet actions can be determined by passing them through several modules that can act on the given packet. This process may mean setting up an action for the flow in the current switch, or sending out flow rules to multiple switches to direct the packet along a given path. This is a simplified view of controller operation, a detailed description of controller application functionality is beyond the context of this paper but is described by~\cite{Astuto2014}.

There are various controller applications currently available, each with specific qualities that may make them more desirable for certain scenarios \cite{fernandez2013comparing}. The NOX OpenFlow controller was the first OpenFlow controller to be developed, because it was built alongside the initial version of OpenFlow at Nicira Networks. NOX provides a fast and stable control platform for OpenFlow. It has been developed using using C++ and has been used in many research projects since its creation. Floodlight is another controller which provides a full featured platform for OpenFlow development. It has been developed in Java and provides a modular system that allows a straightforward framework for developers to include their own modules for SDN control.

In this paper, we utilised the Floodlight controller due to its simple setup with minimal dependencies and modular design, this allowed straightforward customisation and would allow the integration of the migration system into a Floodlight module at a later date. The Floodlight controller in this project was only used as a standard routing and switching controller, it was not adapted or modified since a custom application for the migration process was implemented. This custom application used flow priority to override the flows inserted by Floodlight with its own; using this allowed the functionality of a standard network before and after the migration process.

\subsection{System Applications}
\label{sub:system_applications}
An example application described in this paper is one that reduces the bandwidth that would be consumed when streaming real time, ultra-high definition content from a single source to an increasing number of clients. These clients are distributed across a geographically diverse network that is likely to be at a global scale. An example scenario would be transcoding ultra-high definition content streamed to cinemas or other venues that have different format requirements. The content would need to be processed by a transcoder to meet the needs of each individual client; this transcoder would initially be located close to where the initial client user base was located. As the number of clients increased in another geographical location, it would become more efficient in bandwidth terms to migrate the transcoder closer to the largest user base. This approach would significantly reduce the bandwidth that would be consumed, since the need to send multiple streams at different resolutions would be removed and only a single high quality stream would need to be transmitted over constrained links. Links such as transatlantic cables between the UK and the USA are potential examples that could benefit from this type of optimisation.

Application layer multicasting can be used to reduce the bandwidth consumed on links between the source of the content and the transcoder; this is accomplished by eliminating most of the wasted bandwidth that would otherwise be consumed by repeated transmission across those links \cite{hosseini2007survey}. This technique provides an excellent solution to one of the many issues linked with transmitting ultra-high definition content across a network which has strict bandwidth constraints.

\section{Transcoder Placement Optimisation}
\label{sec:transcoder_placement_optimisation}
In this section we will detail the process of optimising the positions of a set number of transcoders in a network. It will illustrate that there are significant advantages with the use of transcoders to both transcode content into multiple resolutions, along with providing application layer multicasting throughout the network. The placement of these transcoders can influence whether they provide a positive or negative effect on the total network load being placed on the links of the network. This is why it is beneficial to place them in positions that have been optimised to minimise the total traffic in the network. This section is an extension of a previous paper \cite{farrow2013optimising}, which presented the concept of optimising transcoding resources within network.

\subsection{Placement Problem Statement}
\label{sub:problem_statement}
The placement of transcoders will be described over a graph $G(E,V)$, of edges $E$ describing physical links in the network and nodes $V$ representing attachment points for clients, servers and transcoders.  The set of source nodes providing content will be denoted $S \subseteq V$; $T \subseteq V$ is the set of destinations for the content; and, $A \subset V$ is the set of compute resources in the network that are candidates for transcoder placement. In practice each node may host a number of clients, a smaller number of nodes will host sources and transcoders. This scenario assumes that candidate transcoder nodes, $a \in A$, are hosted in an elastic cloud computing technology such as Amazon EC2 giving the possibility to scale up resources as needed, with only cost being the restricting factor.  However, each link, $e \in E$, in the network has a limited transmission capacity $u(e)$.

The traffic travelling through the network from $s$ to $t$ is defined as $R(s, t, b_x)$ where $b_x$ is the bitrate for codec $x$, and where $b_x \in B \left \{\textit{codec bitrate}: HD, 4K, ...\right \}$. These traffic requirements can be stated as a set of demands, $D = \left \{d_{s,t} \;|\; s \in S, t \in T, d_{s,t}=b_x,b_x\in B \right\}$. In practice it may not be possible to transmit all of the demanded traffic because of the capacity constraint and thus the successful demands are defined as $D' \subseteq D$. The problem then becomes to choose a set of paths for the demands, denoted $P(D)$, to meet the objective
\begin{equation}
  \label{eq:maxD}
  \max |D'|  
\end{equation}
subject to:
\begin{equation}
	\sum_{p \in P(D)} p(e) \leq u(e) \quad \forall e \in E
\end{equation}
\begin{equation}
	p(d) \geq d_{s,t} \quad \forall d \in D
\end{equation}
where: $p(e)$ is the traffic on edge $e$ associated by a demand path $p \in P(D')$, the set of all paths for the solution; and, $p(d)$ is the traffic on a path fulfilling demand $d$.

The process described here is stage one of the problem to be solved, it allows for optimisation of the transcoder placement in the network; however if this process is to be used with a dynamic system with optimisation throughout content transmission, it is required to move transcoders while streaming the content. Otherwise the optimisation can only be made before the streaming takes place, which although may be an improvement over random placement or statistical analysis based placement, is not enough to maintain efficiency when dealing with varying client demands during transmission. This is the issue that stage 2 of the problem looks to address and it is detailed in section \ref{sec:system_design}.

\subsection{Algorithmic Solution} 
\label{sub:algorithm_approach}
The problem described in \ref{sub:problem_statement} is known to be NP-complete~\cite{Even76} and thus there is no known polynomial-time optimal solution as testing every single combination of variables requires a time exponential in one or more of the factors that can be varied. Consequently, this paper develops a heuristic algorithm to provide a \emph{good} solution in a reasonable time. The heuristic is compared to a genetic algorithm (GA) which can find good solutions although in a much longer period of time. The metric chosen for comparison is run-time under the condition that the solution quality between the two algorithms is the same (or very close). 

A GA was chosen for comparison because it provides a solution that has a high chance to be very close to the optimal solution, as long as suitable parameters are chosen. Although a GA is relatively slow compared to a heuristic approach, it is considerably quicker at finding good solutions (sometimes optimal) than testing every conceivable solution and can generally get closer to the optimal solution than a heuristic algorithm.
Because of the long run-time of a  GA, it is generally reserved for offline route calculation. However, adequate solutions can be achieved in shorter time by reducing certain solver variables such as the number of generations or chromosomes.

Using a combination of the GA results and analysis of the problem, it was possible to construct a heuristic algorithm that could be used in place of a GA in an online system. This is the main purpose of the first stage of this papers work and will be considered after briefly describing the design of the GA. In both cases the algorithms were constructed in R to make sure that the results were comparable in terms of run-time.

\subsection{Genetic Algorithm Design} 
\label{sub:genetic_algorithm_design}
The GA was implemented with the ability to perform both chromosome crossover as well as mutation functions. This implementation provided a reasonable solution when applying the functions over multiple generations; the number of generations was optimised to give both an optimal solution and as short as possible running time. This optimisation was determined by producing results for varying generation counts and then analysing the values to find the point at which adding more generations did not provide any significant improvements in the solution.

To produce the most optimal solution without significantly sacrificing the speed of computation, the ideal values for the GA were: 10 Generations, Population of 50, crossing and mutating 50\% of the chromosomes at random each generation. To simplify the GA crossover function, it was decided to remove chromosomes with errors after crossing rather than apply a repair function; it is important to note that this was implemented to reduce the delays in computation that would be caused if a large number of chromosomes needed to be altered. 

\subsection{Heuristic Algorithm Design} 
\label{sub:heuristic_algorithm_design}
The heuristic was developed through the use of both a mathematical approach as well as testing; these methods provided a heuristic which performs almost as well and sometimes better than the GA, but produces a solution in considerably less time.

The heuristic looks at the available locations for transcoders and scores them using a fitness function, the best location is then picked as a starting location. The fitness function uses a combination of Dijkstra's shortest path algorithm, a combination of the load and codec used for the demands and also the number of connected links available to the transcoder location. The locations of the other transcoders are then chosen by selecting locations that are a certain separation from current transcoders and then running a group scoring function. Reducing the number of suitable transcoder locations in this way provides relaxed selection conditions for the algorithm, which accelerates the selection process.

This approach is much faster than a brute force approach of scoring every combination, but it still provides an acceptable solution for the given problem. By keeping the locations of the transcoders separated as much as possible, it provides a high probability that there will be a transcoder at an acceptable distance from each client. Keeping the transcoders as separated as possible also reduces the possibility of a transcoder being overwhelmed by clients while other transcoders remain unused; this is true for networks with a reasonably uniform distribution of clients.

The basic design of the heuristic is shown as algorithm \ref{alg:heuristic}, which details most of the process. It should be noted that the \textsc{Dijkstra} function shown in the algorithm uses Dijkstra's shortest path algorithm to return the hop distance from the given nodes in graph $G$, while $\lambda$ represents the separation constant that is used to optimise the algorithm for speed or accuracy; this separation constant is explained in more detail in~\ref{sub:methodology}. Additionally, the \textsc{degree} function used in algorithm \ref{alg:heuristic} returns the number of active links available on the given attachment point. The group scoring function (\textsc{Score}) performs similar actions to the scoring mechanism shown from lines 4-14 in algorithm \ref{alg:heuristic}, but instead of using all the possible transcoder locations the closest transcoder to each client from the given group is used for the scoring function; this is performed across all the client requests with their scores being added together to generate the final group score. The \textsc{findClosestTranscoder} function used in \textsc{Score} uses Dijkstra's shortest path algorithm to find the closest transcoder to the client and then it returns the location of this transcoder.

\begin{algorithm}
	\caption{Transcoder placement for up to $N$ transcoders, with separation constant $\lambda$}
	\label{alg:heuristic}
	\begin{algorithmic}[1]
        \State $minScore \gets \infty$  
        \State $transcoder \gets null$    \Comment{a transcoder location}
        \State $transcoders \gets null$  \Comment{a list}
		\For {$a \in A$}
			\State $score \gets 0$
			\For {$t \in T$} \Comment{for all destinations $T$}
		    	\State $dist\gets \Call{Dijkstra}{G,a,t}$
		    	\State $score\, + = (dist \times d_{s,t})$                  
		    \EndFor
		    $score \gets score / \Call{degree}{a}$
		    \If {$score < minScore$} 
            	\State $minScore \gets score$
            	\State $transcoder \gets a$
            \EndIf
		\EndFor		
		\State $transcoders \gets \Call{append}{transcoders,transcoder}$
		\If {$N > 1$}			
			\State $sepDist \gets \lambda \times |V|$
			\For {$1 \ldots (N-1)$}
				\State $suitLoc \gets null$ \Comment{list of suitable locations}
				\While {$suitLoc == null$}
					\For {$a \in A$}
						\For {$k \in transcoders$}
							\If {$sepDist < \Call{Dijkstra}{G,a,k}$}
								\State $suitLoc \gets \Call{append}{suitLoc,a} $
							\EndIf
						\EndFor
					\EndFor
					\State $sepDist - 1$
				\EndWhile
				\State $grpScores \gets \infty$ \Comment{list, initially length 1}
				\State $bestLoc \gets null$
				\ForAll {$j \in suitLoc$}
					\State $testGroup \gets \Call{append}{transcoders, j}$
					\State $gScore \gets \Call{Score}{testGroup}$
					\If {$gScore < \Call{min}{grpScores}$}
						\State $bestLoc \gets j$
					\EndIf
					\State $grpScores \gets \Call{append}{grpScores,gScore}$
				\EndFor		
				\State $transcoders \gets \Call{append}{transcoders,bestLoc}$			
			\EndFor
		\EndIf		
		\State\Return $transcoders$
                \item[]
		\Procedure{Score}{$locations$}
			\State $groupScore \gets 0$        
			\For {$t \in T$}
				\State $trans = \Call{findClosestTranscoder}{t}$
		    	\State $dist\gets \Call{Dijkstra}{G,trans,t}$
		    	\State $groupScore\, + = (dist \times d_{s,t})$  		    	
		    \EndFor
		    \State\Return $groupScore$
		\EndProcedure
	\end{algorithmic}
\end{algorithm}

\subsection{Methodology}
\label{sub:methodology}
There are three parameters that this stage of the work measures: the run-time of the heuristic compared to the GA; the comparison of the solution quality which is the value of the objective function~(\ref{eq:maxD}); and also the solution quality relating to the reduction in network load that is accomplished in certain scenarios, especially when concerning non-blocking environments. To allow a fair comparison between the GA and heuristic algorithm, the GA run-time was measured as either: the time to achieve the same or better score in the fitness function than the heuristic; or, when the stop condition is reached. The latter is to allow for the fact that sometimes (rarely) the GA does not reach as good a solution as the heuristic. To test the quality of the heuristic compared to the GA, the GA was run for a longer period after it matched the heuristic objective function result to see if it could improve on this value.

The performance optimisation of the GA was briefly described in~\ref{sub:genetic_algorithm_design}, as is standard procedure for determining GA operating parameters. The heuristic algorithm required more detailed specific parameterisation which is described here. The algorithm has been designed such that it only requires one parameter to be adjusted: specifically the separation parameter introduced in~\ref{sub:heuristic_algorithm_design}. It was found that this parameter affects both the run-time and quality of the algorithm.

The separation value can be set to a range of values that theoretically fall between 0 and 1, however selecting a value over 0.1 will provide in almost all cases the same solution as 0.1 but with increased runtime; this presents a practical range of values between 0 and 0.1. It should be noted that 0.1 was found to be a suitable maximum value after testing multiple values across various network scenarios, the results were then analysed to identify the point at which no significant increase in performance was achieved. The exact use of the separation constant can be seen in Algorithm \ref{alg:heuristic}, which shows how it is used along with the number of network nodes in the network to produce a separation distance; this ensures performance is maintained as the network increases in size. Transcoders that are at least the separation distance away in hop count from current transcoder locations are selected for scoring. From this we can ascertain that a smaller separation distance (i.e. a smaller separation constant) will generally produce a larger selection of transcoders that meet the distance requirement, with a larger distance (i.e. larger separation constant) producing the opposite effect. Therefore we can assume that the larger pool of suitable transcoder locations will take longer to score than the smaller group, however there is a greater probability of finding a better quality transcoder location in the larger group. Using this information it is clear that we can configure the algorithm using the separation value for either a fast solution output or a better quality output.

There are a number of scenario parameters that affect the performance of the algorithm and that are being investigated. One of the most important of these scenario parameters is one that relates to the ability for the algorithm to scale with the size of the network. We would expect that their performance should not drastically diminish as the number of nodes increases. Other network factors could also influence the algorithm performance such as the proliferation of data centres that a transcoder could be migrated to, as well as the number of transcoders that will be up and running in different geographical locations at any given time. In the experiments this is tested by increasing the network size.

It is also important to note that the number of clients and distinct videos will be a large determining factor to the algorithms performances. Having a low number of clients or high number of distinct videos can reduce the effectiveness of application layer multicasting at the transcoders; this reduction of performance is attributed to the observation that for application layer multicasting to work, a distinct video will need to be streamed to more than one client from a single transcoder for the bandwidth savings to have an effect. The effect of this has been tested by varying the number of clients within the network.

\subsection{Optimisation Results}
\label{sub:optimisation_results}
In cases where separation values are different we expect to get different solution qualities and run-times. As the GA run-time is quantified by reaching the same result (or better) as the heuristic, the GA run-time will depend upon the separation value used in the heuristic. Therefore, the GA is denoted with a separation value on the graphs so it can be compared with the heuristic of the appropriate separation value, even though the GA does not explicitly use the separation parameter.

\begin{figure}[ht]
	\centering
	\subfigure[Blocking Reduction.]{
		\adjincludegraphics[width=\columnwidth,keepaspectratio]{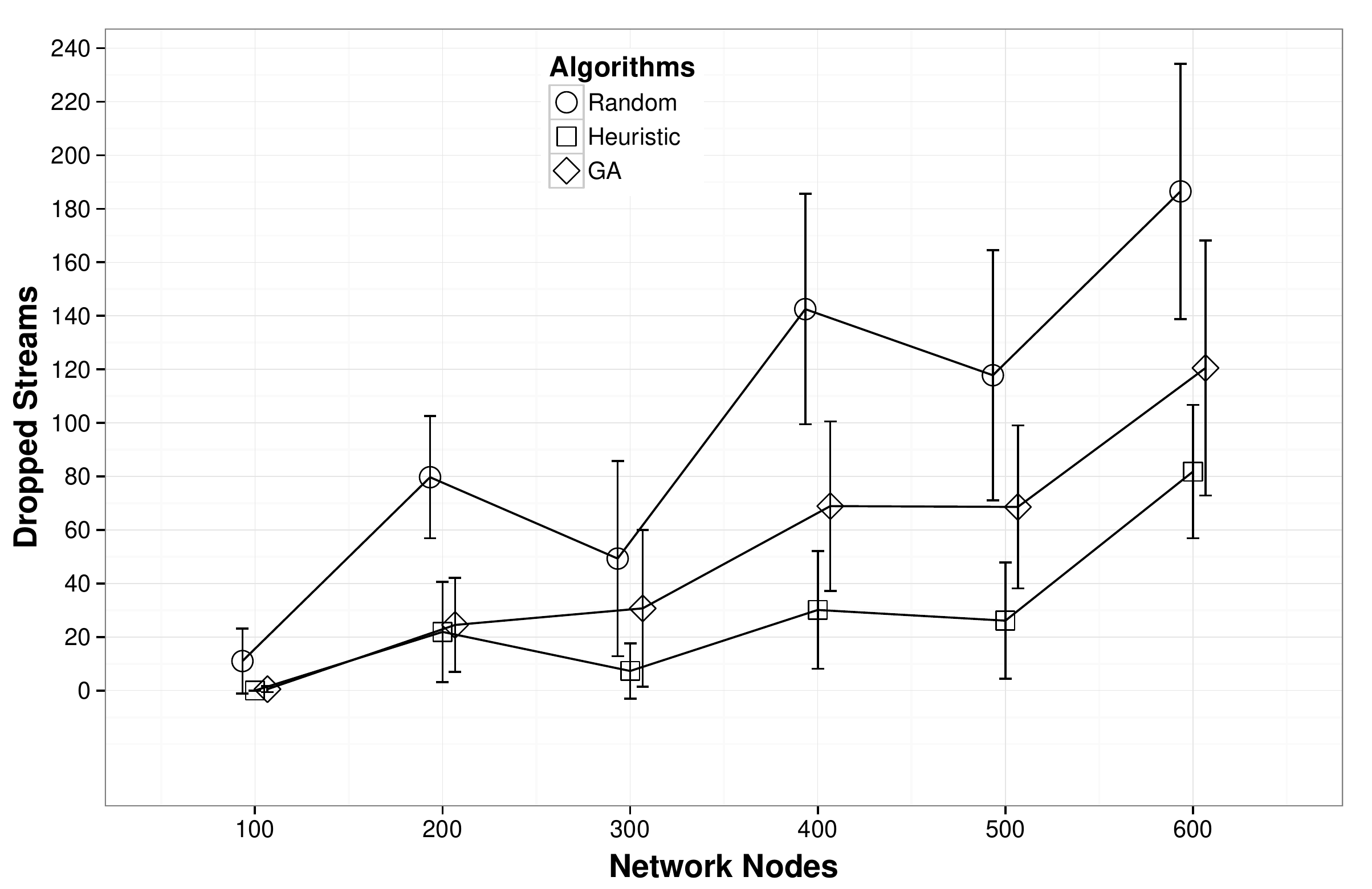}
		\label{fig:blocking_load}
	}
	\subfigure[Run-time performance.]{
		\centering
		\adjincludegraphics[width=\columnwidth,keepaspectratio]{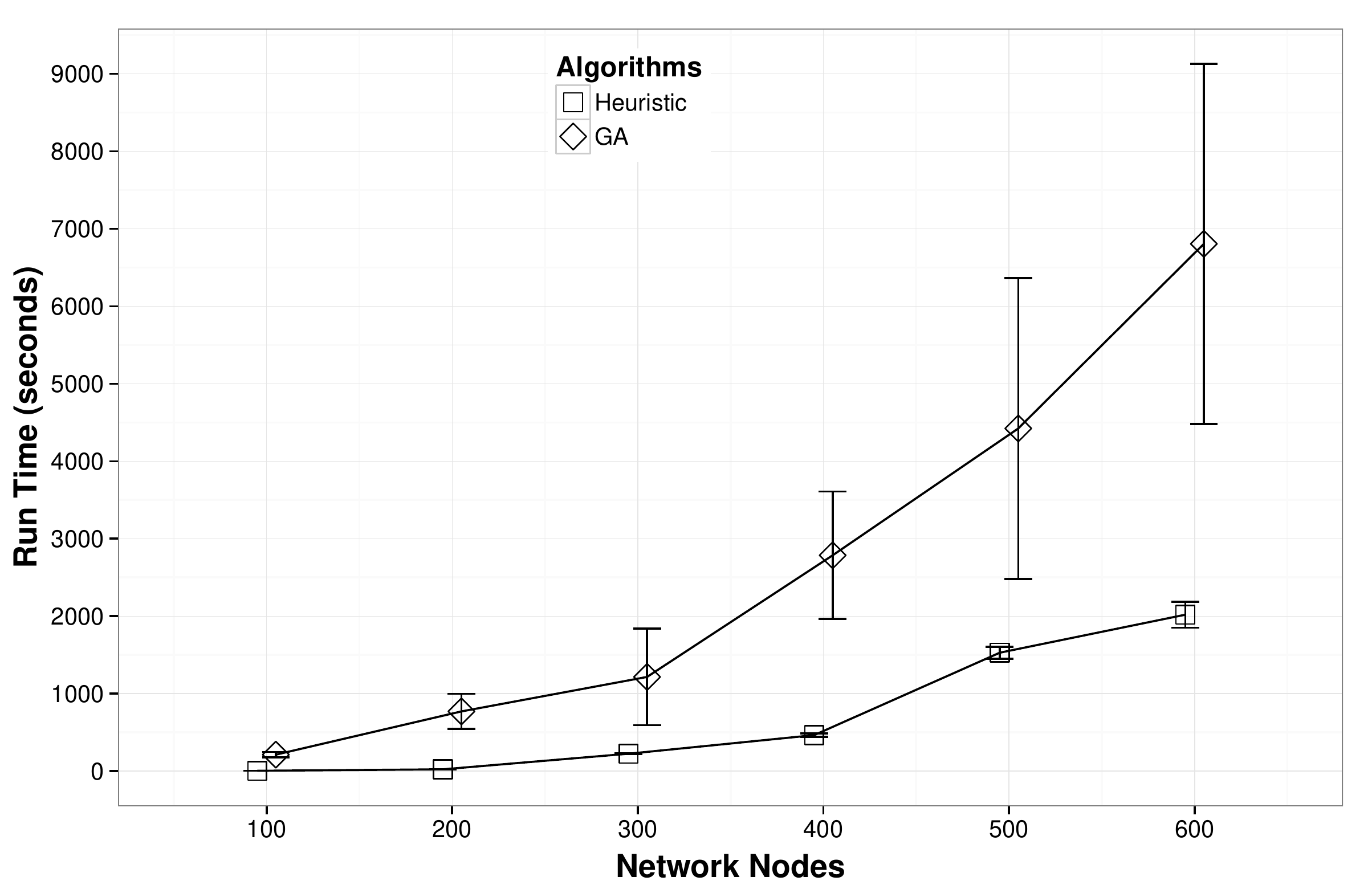}	
		\label{fig:blocking_time}
	}
	\caption{Comparing (a) the quality of the solution at reducing blocking in the network and (b) the run-time for different network sizes. Run-time is omitted for random placement due to the negligible time factor.}
    \label{fig:blocking_results}
\end{figure}

Apart from the results shown in Fig. \ref{fig:blocking_results}, all results were gathered in a non-blocking environment. For the cases where it is non-blocking, in terms of comparing network quality, the objective function in (\ref{eq:maxD}) is instead expressed as the total traffic admitted to the network, termed \emph{network load}, where the objective is for this parameter to be as low as possible to allow for further demands to be met; this reduction in traffic translates to being able to fulfil a larger number of demands, because of the increased available bandwidth in the network.  Formally, we define network load $L$ as:
\begin{equation}
  \label{eq:1}
  L = \sum_{d \in D} d_{s,t} |E(d_{s,t})|
\end{equation}
where $d_{s,t}$ is the bitrate of the video demand $d$, defined earlier, and $E(d_{s,t})$ is the set of edges that $d$ is transmitted over.

\begin{figure}[t]
	\centering
	\adjincludegraphics[width=\columnwidth,keepaspectratio]{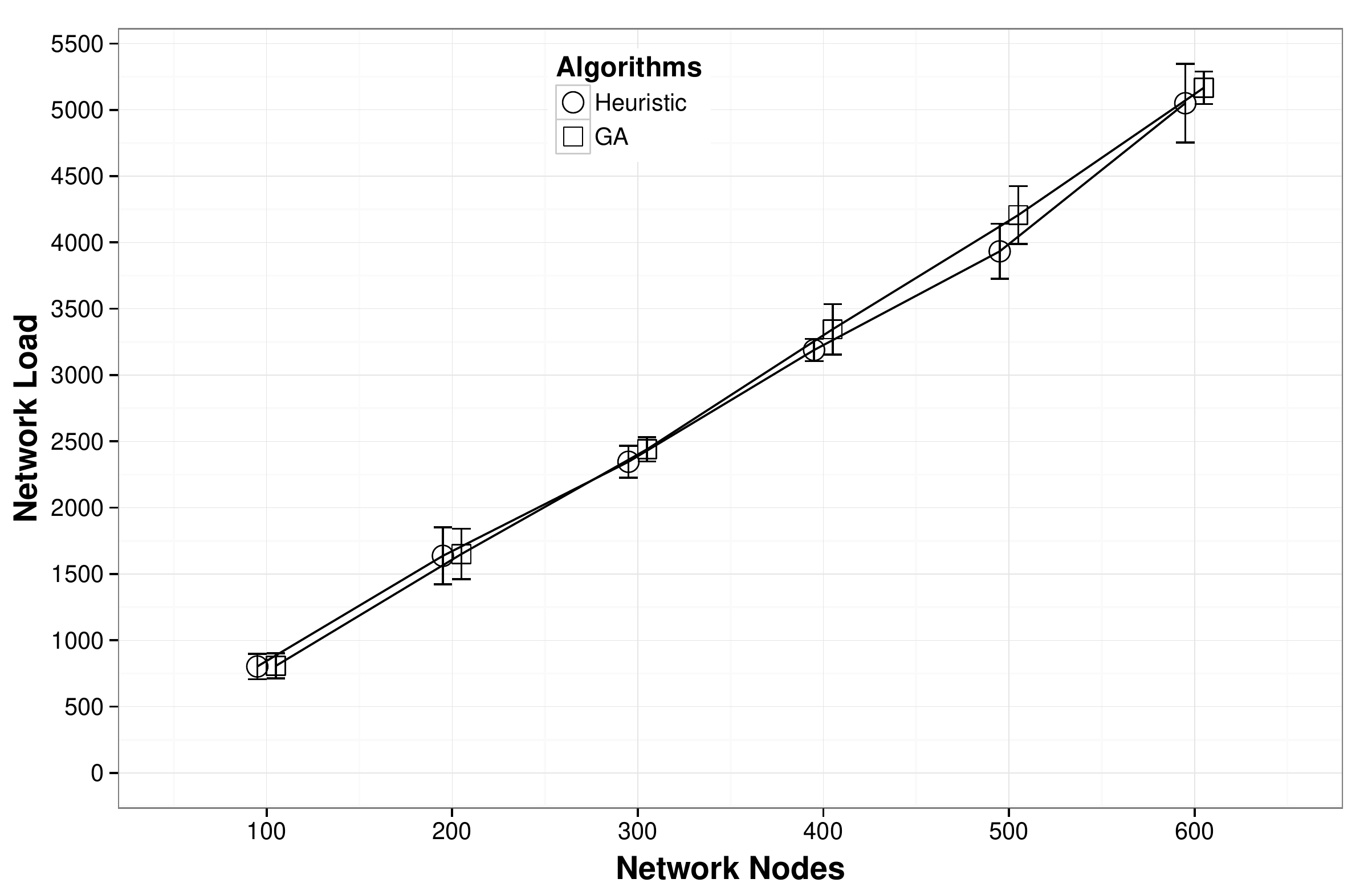}
	\caption{Admitted network load for the heuristic and GA where the GA was continued without stopping when it reached the same result as the heuristic. Notably the GA produces highly compatible results.}
	\label{fig:ga_no_stop}
\end{figure}

It is also important to note that apart from Fig. \ref{fig:ga_no_stop}, the GA was given the stopping criteria of reaching a result similar to that of the heuristic, so that the time analysis would be a fair comparison. But even without the stopping criteria, it is shown in Fig. \ref{fig:ga_no_stop} that the GA does not provide a significantly improved solution even when given the large amount of time required to finish its calculation; this was reasoned as an acceptable result based on this fact. 

\begin{figure}[t]
	\centering
	\adjincludegraphics[width=\columnwidth,keepaspectratio]{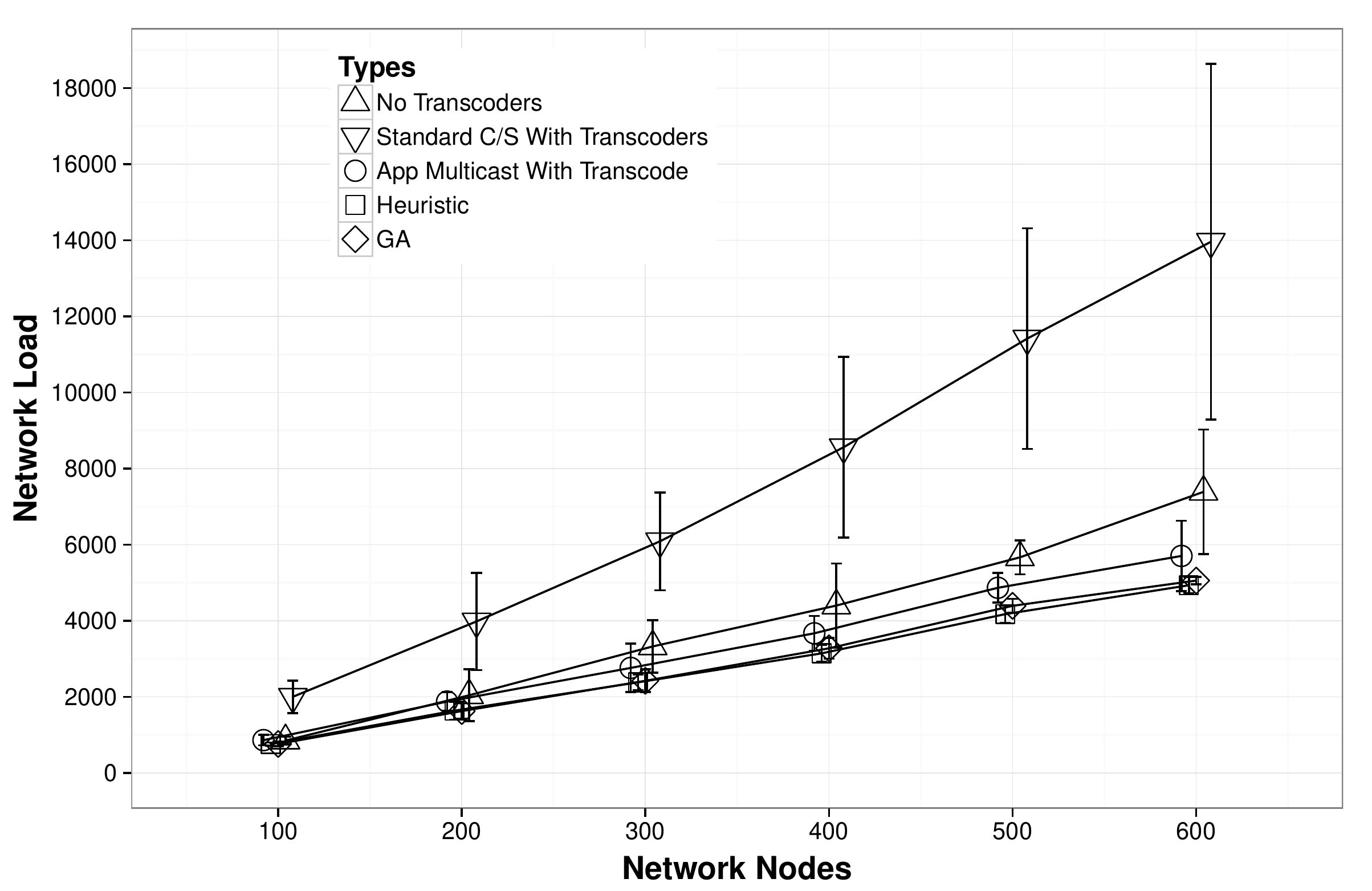}
	\caption{Comparison of network load under a non-blocking scenario showing the benefit of using transcoders to reduce overall traffic load.}
	\label{fig:no_blocking_or_transcoders}
\end{figure}

It is also notable to mention the importance of using transcoders to both transcode content to other formats as well as using them to perform application layer multicasting. Fig. \ref{fig:no_blocking_or_transcoders} shows the effects of transcoder usage in a non-blocking environment; it is clear from these results that using transcoders provides an advantage in reducing traffic load in the network. If these results were collected in a constricted network which allowed blocking, it would be expected that an increased amount of blocking would occur compared to scenarios that used transcoders.

Fig. \ref{fig:blocking_results} presents the results given from the objective function (\ref{eq:maxD}), while using the best case scenario of the heuristic with 0.01 as the separation value. From these results it can be seen that the heuristic considerably improves on the GA solution, while remaining significantly faster at achieving this result. The random placement is included for comparison purposes only. It should be noted however that for networks above 700 nodes the performance of the heuristic does begin to diminish in quality, although not to the point that brings it below that of random placement. The number of nodes reflects switching and attachment points for devices, which includes servers, transcoders and clients. Thus, a network size of 600 is already large and may support a very large number of client devices.

\begin{figure}[ht]
	\centering
	\subfigure[Varied separation value network load performance.]{
		\adjincludegraphics[width=\columnwidth,keepaspectratio]{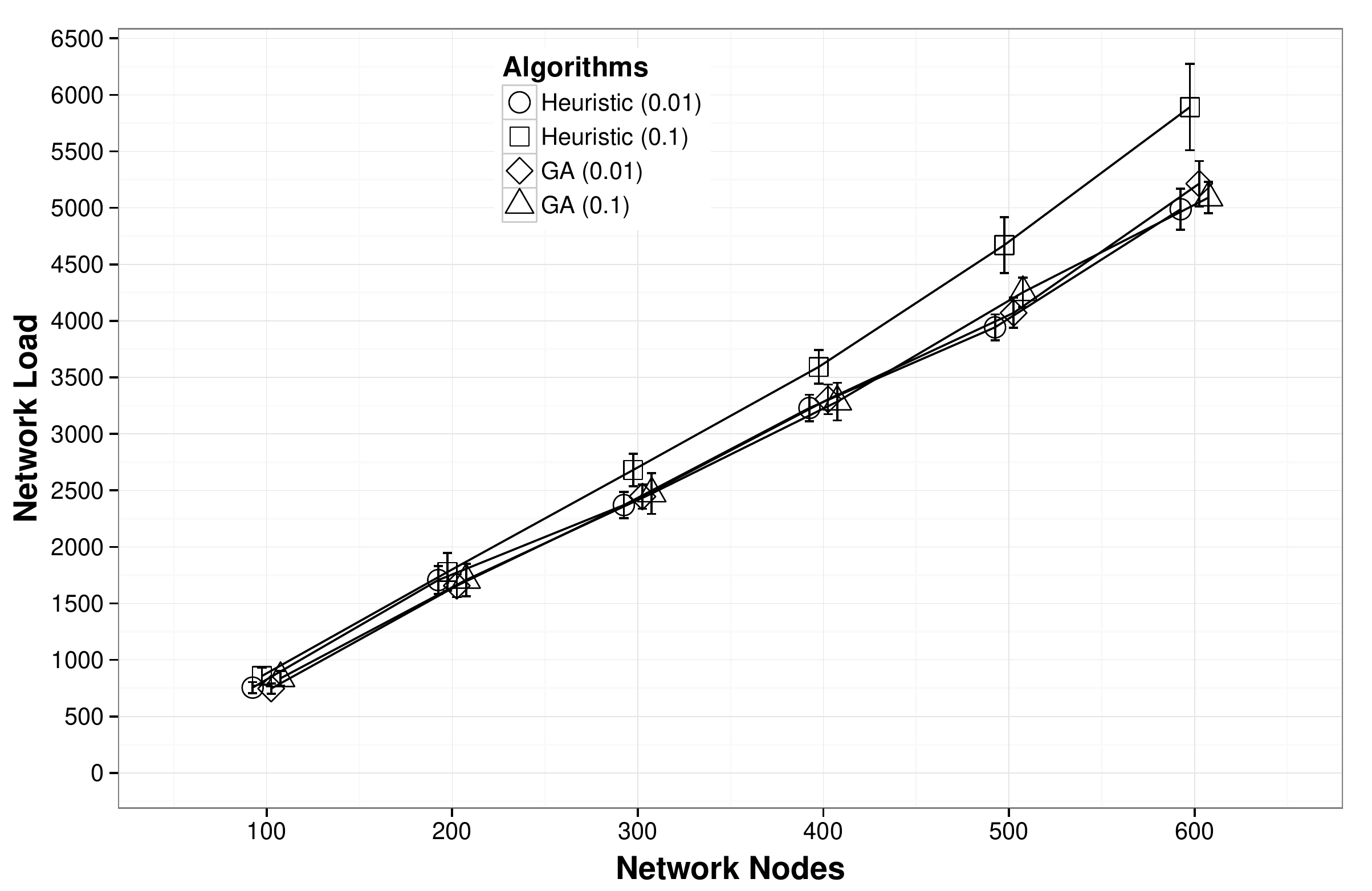}
		\label{fig:seperation_load}
	}
	\subfigure[Varied separation value run-time performance.]{
		\centering
		\adjincludegraphics[width=\columnwidth,keepaspectratio]{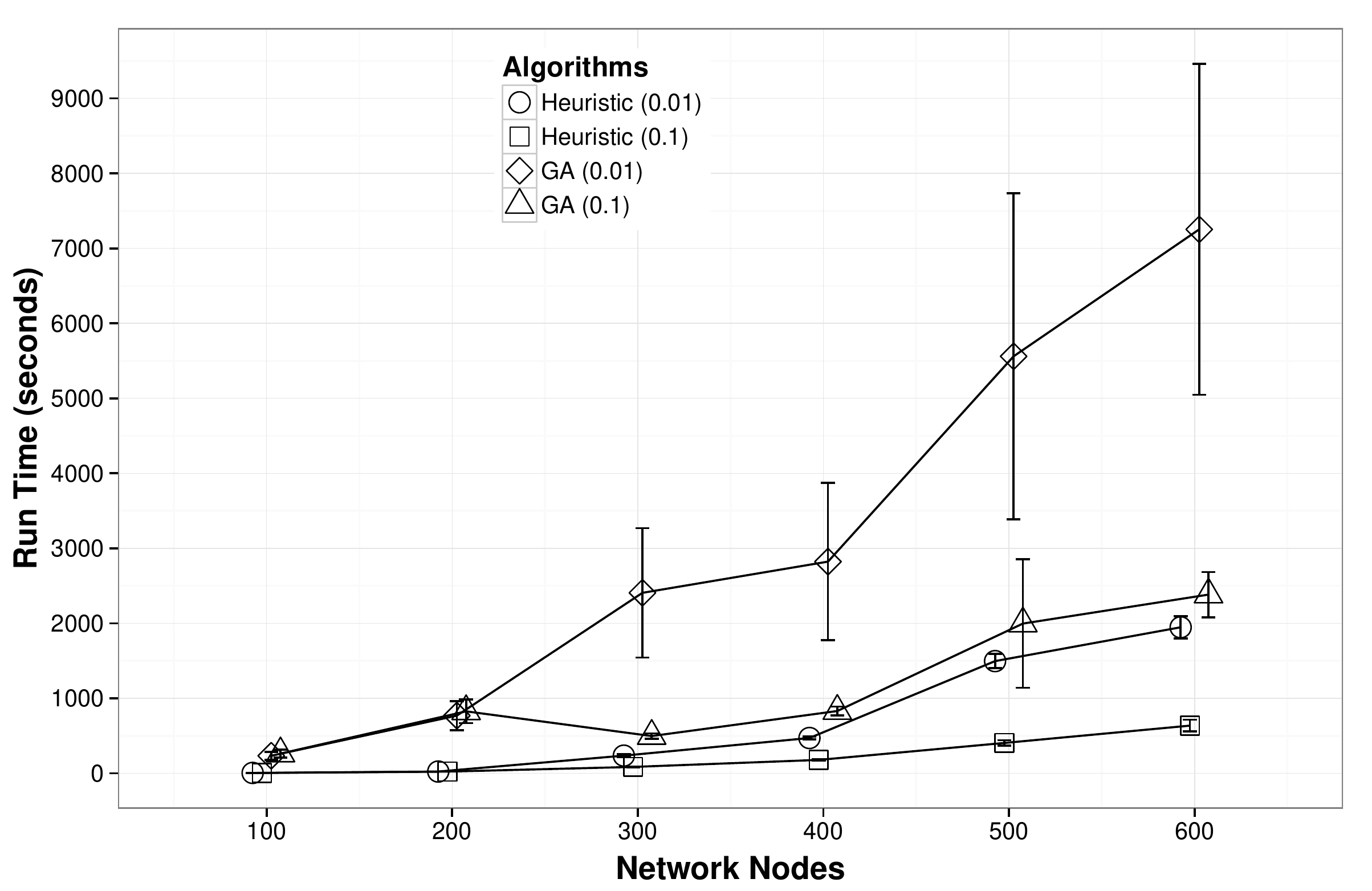}	
		\label{fig:seperation_time}
	}
	\caption{Comparing (a) the quality of the solution and (b) the run-time for different separation values and network sizes. Note the run of the comparable GA is denoted with the separation value for identification purposes only.}
    \label{fig:separation_results}
\end{figure}

Fig. \ref{fig:separation_results} shows how the separation parameter can effect the outcome of the solution quality, as well as the run-time of the heuristic algorithm. The smaller separation value provides a better solution, however it takes longer to achieve this; in contrast the lower separation value provides a slightly worse solution in a shorter period of time, however there is only a marginal difference in solution quality and certain scenarios would find this more than adequate for the improvement in run-time. It should be noted that separation values other than 0.01 and 0.1 were used during testing, the two values presented in the results, shown in Fig. \ref{fig:separation_results}, were chosen to give a clear example of the how the balance between solution quality and run-time can be influenced. Values between these points produce results that can be expected, lying between the results of 0.01 and 0.1. As discussed in~\ref{sub:methodology}, using values above 0.1 does not introduce an improvement in solution quality or run-time, so values above this limit were omitted. 

\begin{figure}[ht]
	\centering
	\subfigure[Varied client numbers network load performance.]{
		\adjincludegraphics[width=\columnwidth,keepaspectratio]{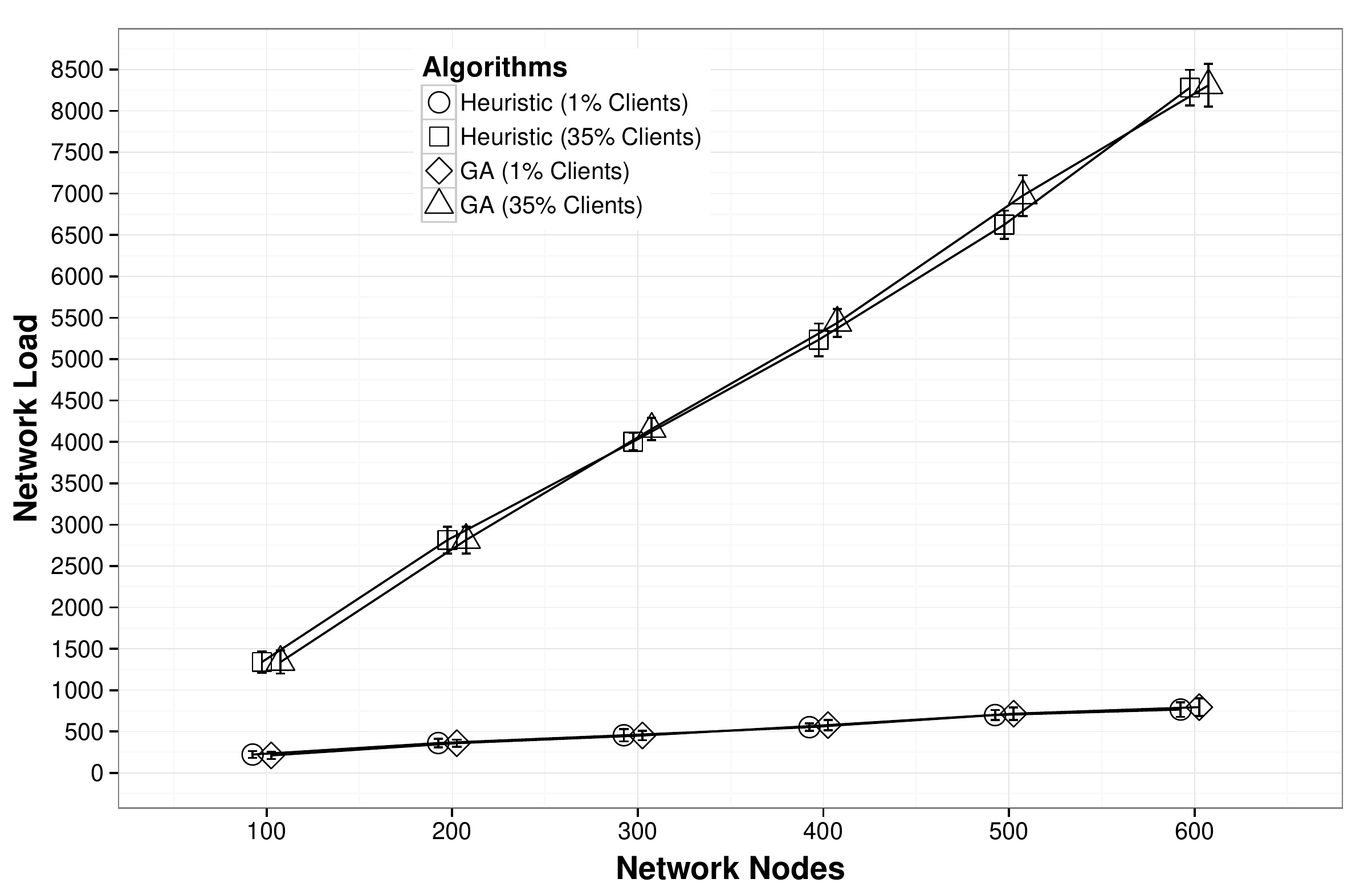}
		\label{fig:clients_load}
	}
	\subfigure[Varied client numbers run-time performance.]{
		\centering
		\adjincludegraphics[width=\columnwidth,keepaspectratio]{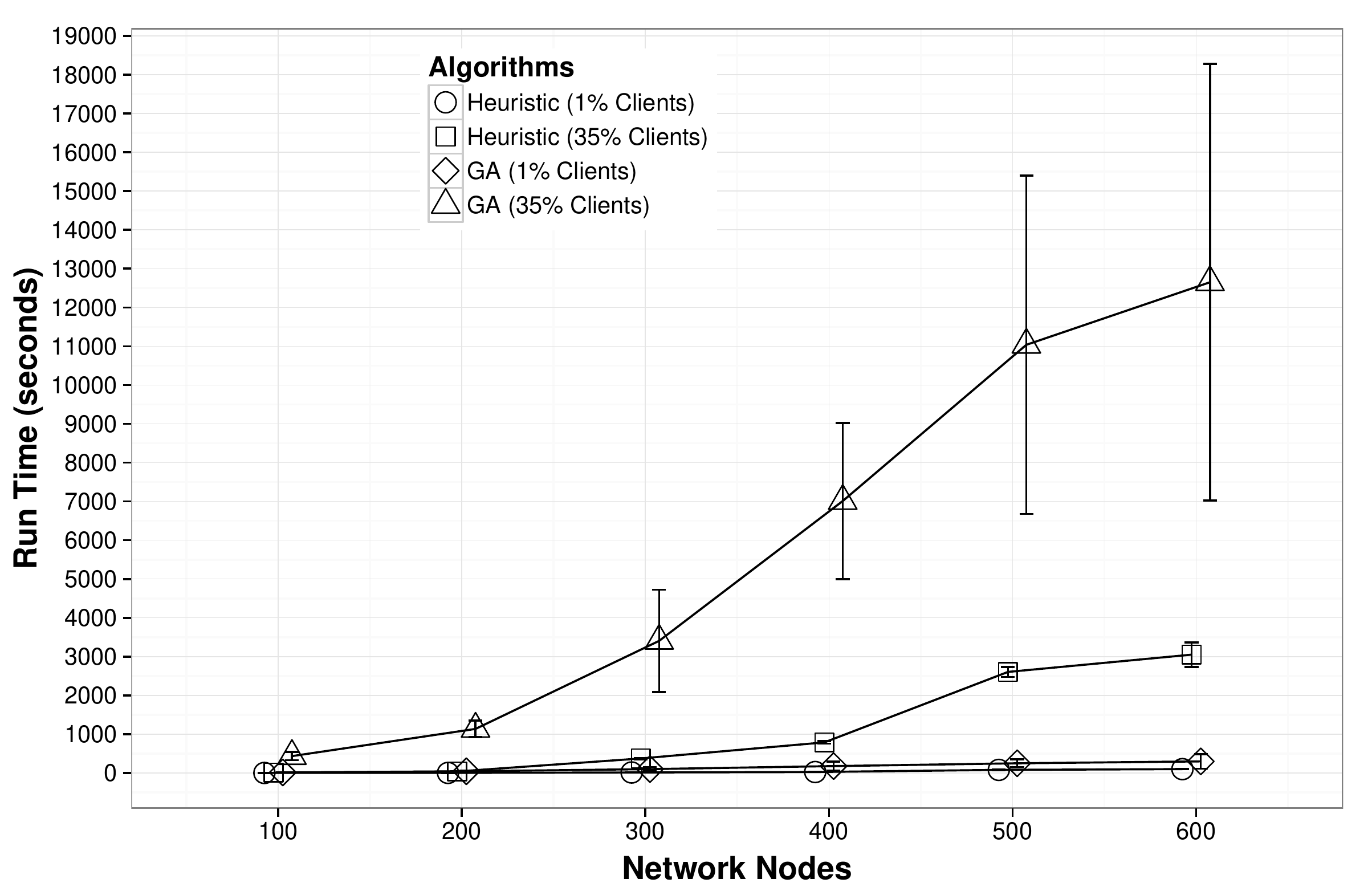}	
		\label{fig:clients_time}
	}
	\caption{Comparing (a) the quality of the solution and (b) the run-time for different numbers of clients (1\% or 35\% of total nodes hosted clients) and for different network sizes. Results for values between 1\% and 35\% are omitted for clarity}
    \label{fig:client_results}
\end{figure}

From Fig. \ref{fig:client_results} we can determine that the solution quality of the heuristic is maintained throughout the variation in client numbers; along with the run-time being consistently less than that of the GA when client numbers begin to rise. Furthermore, it should be noted that client values between 1\% and 35\% were also tested, but they were omitted from Fig. \ref{fig:client_results} for clarity; however, it is important to highlight that the results collected between these values show an expected gradual increase as client numbers rise, in line with the presented results.

From the results shown in Figures \ref{fig:ga_no_stop}, \ref{fig:separation_results}, \ref{fig:client_results}, it can be seen that the heuristic algorithm performed as well, or better, than the GA; with respect to network load, the heuristic also on occasion produced a better solution than the GA. In all cases the heuristic obtained a result that was within 16\% of the GA network load value. More importantly than this, the heuristic is significantly faster at providing a solution than the GA, which means it could be utilised in an online system for optimising current traffic in a network. This is an important factor based on the given scenario of using this system to dynamically migrate transcoders based on client demands. 

\section{System Design}
\label{sec:system_design}
In this section of the paper, we present the migration system which can migrate a transcoder during a live transmission with minimal interruption to the receiving client. This mechanism would be utilised after the algorithm in \ref{sec:transcoder_placement_optimisation} determined that it would be beneficial for the transcoding to be moved to another location. This section will also provide the specific details of the test setup as well as presenting collected results to show the advantages of using this system over current alternatives.

The reduction in the delay between packets arriving from the old transcoder and the new transcoder is the main metric that this system aims to improve, as this reduction improved the viewing quality for the client during the migration process. For this reason, the delay between these two streams was the focus for the collection and analysis of results; other factors were also recorded for analysis, but were not essential for the migration to be deemed as a success.

\subsection{System Architecture}
\label{sub:system_architecture}
The system architecture implemented for this paper, uses a single client receiving video content from a server. This is a simplified example of the proposed scenario, which is presented as a system to provide content to multiple clients at varying resolutions and bitrates. This simplified example was used due to restrictions to the equipment available for collecting results, as well as time restrictions in completing the project. Initial results however have shown that the system functions successfully with multiple clients requesting the same content at different resolutions; in this scenario, the transcoder receives a single stream at the highest requested resolution and then uses application layer multicasting to distribute the content at multiple resolutions to the respective clients. The experiments used the FFmpeg transcoder~\cite{ffmpeg}. 

It is important to note that the link between the two switches was set up to produce varying levels of delay to provide accurate result set. The values for these were used based on the HPDMnet global testbed network \cite{Mambretti2011} that was in place and being utilised for testing. The link between PSNC and Starlight has a round trip delay of 125ms, which presented a useful metric to test the system with. A 250ms delay was also tested to provide an insight in the performance of the system on high latency systems. These values were also compared against a link with negligible latency ($<1$ms), to show how the system is not adversely affected by the link delay. This was an important constraint of the system to focus on, since the content being streamed is envisioned as real-time content, which is also heavily influenced by network latency.

Fig. \ref{fig:migration_stages} shows the system architecture used in testing.

\subsection{Control Software}
\label{sub:control_software}
The control software being used for the test interacted directly with the Open vSwitch SDN switches to set up flows. In a production system it would be integrated directly into the OpenFlow controller, such as the Floodlight controller used in the experiments. The transcoder migration process is initiated as a result of a change in transcoder placement required by the optimisation algorithm that was presented earlier in \ref{sec:transcoder_placement_optimisation}.

\subsection{Migration Process} 
\label{sub:migration_process}
The migration process has many steps to ensure a near instant switchover of streams, this is required to reduce the interruption in the playback of the stream at the client. Although playback systems can handle some interruption/delay in the stream, if it extends more than the playback buffer, the stream will stop and require a longer time to be re-established. This seems due to the fact that after a set time, playback systems tend to treat the delayed stream as a new transmission which requires content analysis to ensure correct playback.

The basic premise behind the OpenFlow migration is that it both replicates the stream at the switches, while redirecting and rewriting packet headers to enable two transcoders to be actively transcoding in parallel with only a single stream reaching the client. The parallel design is essential to allow for the server to establish the new transcoders MAC address in its address resolution protocol (ARP) cache table, without affecting the content stream to the original transcoder which is serving the client. Once the new MAC is established, the old transcoder can be disabled while simultaneously enabling the stream from the new transcoder, to enable a near seamless switchover.

\begin{figure}[ht]
	\centering
	\subfigure[Before Migration]{
		\centering
		\includegraphics[trim=2.5cm 10.4cm 7cm 0.6cm, clip, width=\columnwidth,keepaspectratio]{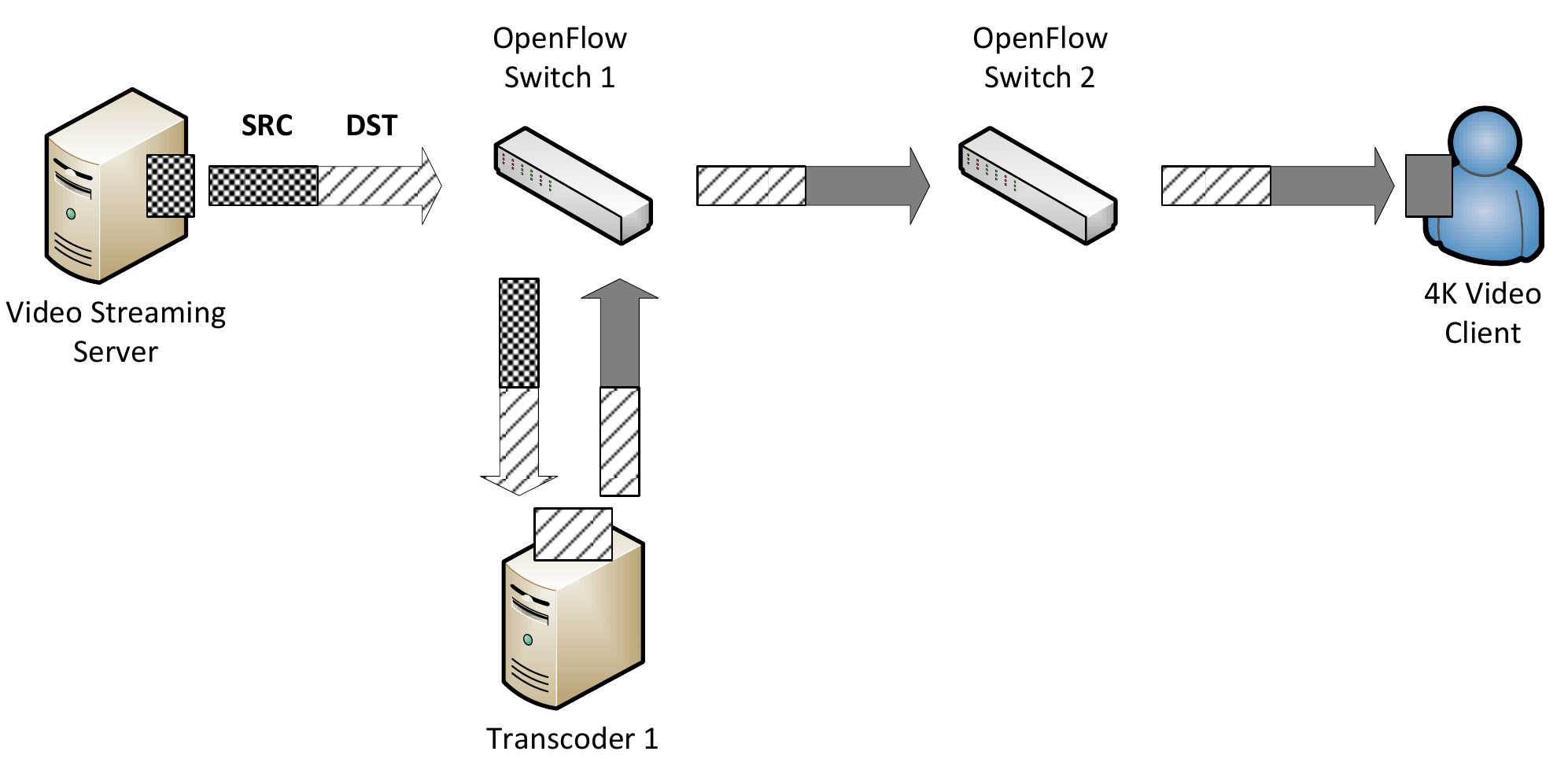}
		\label{fig:stage1_traffic}
	}
	\subfigure[During Migration]{
		\centering
		\includegraphics [trim=2.5cm 10.4cm 7cm 0.6cm, clip, width=\columnwidth,keepaspectratio]{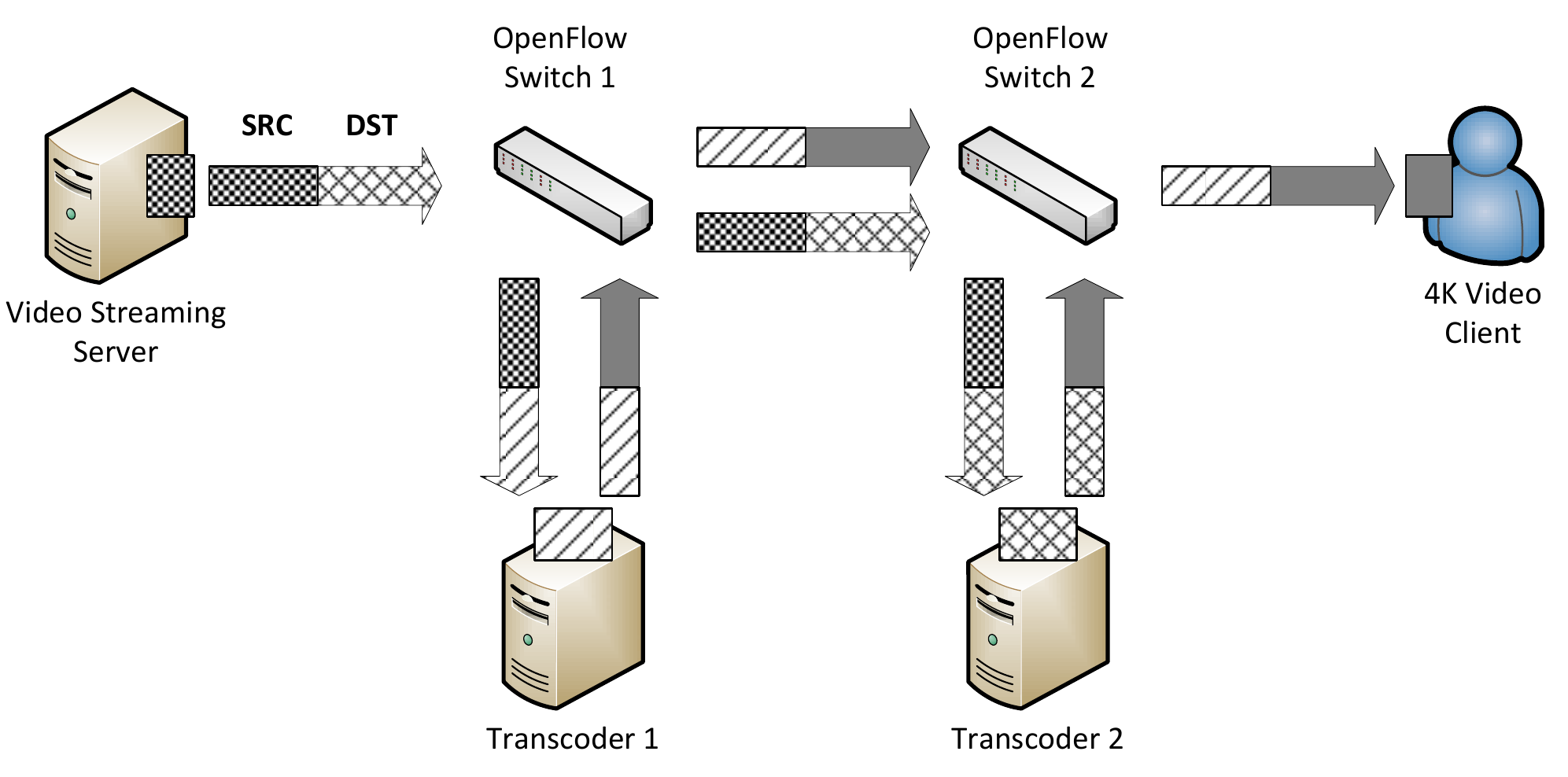}
		\label{fig:stage2_traffic}
	}
	\subfigure[After Migration]{
		\centering
		\includegraphics[trim=2.5cm 10.4cm 7cm 0.6cm, width=\columnwidth,keepaspectratio]{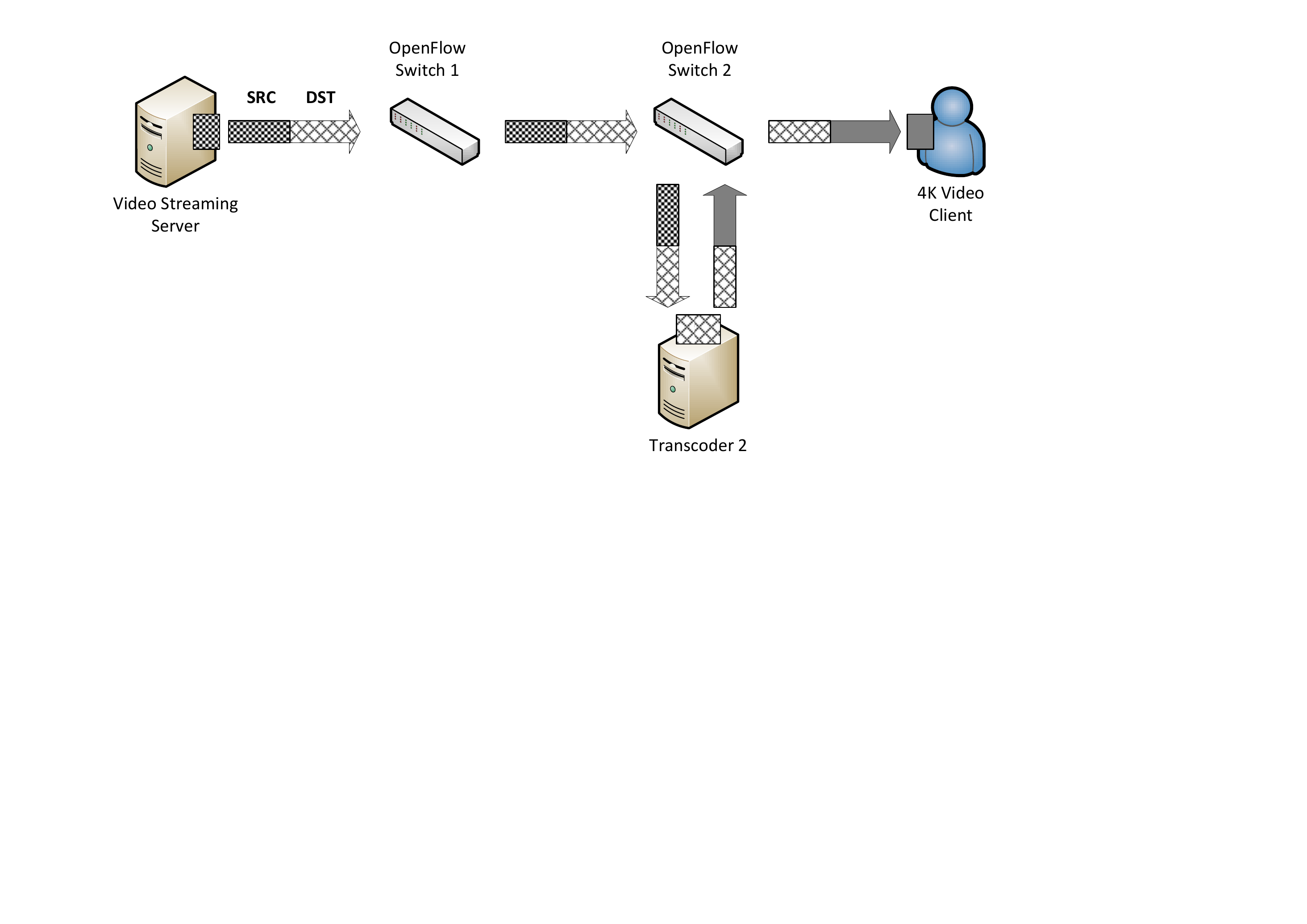}
		\label{fig:stage3_traffic}
	}
	\caption{Migration Stages.}
    \label{fig:migration_stages}
\end{figure}

Figures \ref{fig:stage1_traffic}, \ref{fig:stage2_traffic} and \ref{fig:stage3_traffic} show the simplified concept of the transcoder migration scenario. It is important to note, there are many sub stages involved between these stages, which aid the near seamless migration.

\begin{figure}[t]
	\centering
	\adjincludegraphics[trim={10.8cm 5cm 12cm 1cm}, clip, height={.5\textheight},keepaspectratio]{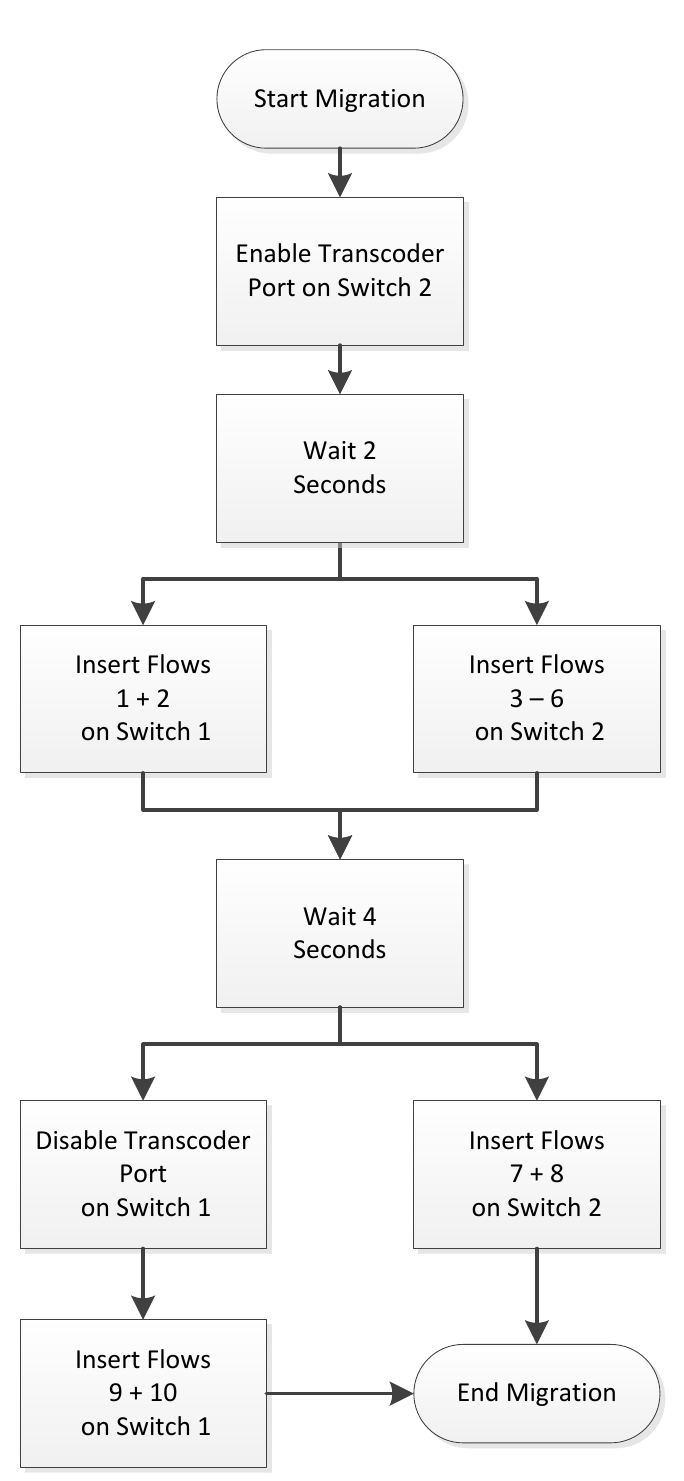}
	\caption{Migration Flow Diagram.}
	\label{fig:flow_process}
\end{figure}

Fig. \ref{fig:flow_process} shows the interaction with two switches during the migration process. The flows referenced in the diagram can be seen in Table \ref{table:flow_table} in appendix A.

The wait times presented in Fig. \ref{fig:flow_process} were chosen as they presented the best performance in the current system, but can be adapted as necessary for specific scenarios. These wait-times are not critical to the overall performance of the system, which is instead determined by the gap in transmission at the receiver which should be as small as possible.  The first wait time is less crucial than the second, since the migration process can actually continue without the port open at that exact step in the migration process; however, the system reliability and performance is increased if this wait time is introduced, as it allows the second transcoder to establish its network connection fully before being sent the data heavy video content. The second wait time is more crucial to the system, as it allows transcoder 2 to begin its transcoding task which involves populating the receiving buffer and analysing any time-dependant factors such as motion vectors. This buffering period can be adjusted in the transcoder, but if it is set too low, there will be issues identifying the content and it will fail to process the video stream. For this reason the buffering time was left to its default setting in the transcoder and the wait time is used to obscure the start up delay in the transcoding process. It should also be noted that the second wait time also provides times to allow the server ARP table to be updated with the new transcoder MAC address, so that by the time of switchover the correct MAC address is already being used by the server; this enables a smooth transition to the new transcoder and prevents any packet header modification from being required after migration completes.

A description of Fig. \ref{fig:flow_process} is as follows:
\begin{enumerate}
	\item Enable both transcoder 2 and the port that it is connected to on Switch 2.
	\item Wait a set time for the port to enable.
	\item Then in parallel:
		\begin{itemize}
			\item On Switch 1:
			\begin{itemize}
				\item Duplicate the stream from the server, sending the unedited stream across to the second switch, while sending another to transcoder 1 but with the MAC destination field rewritten to the transcoder 1 MAC.
				\item Send ARP requests for the transcoder IP to Switch 2.
			\end{itemize}
			\item On Switch 2:
			\begin{itemize}
				\item Send packets destined for the transcoder 1 MAC to transcoder 2 with its MAC destination field set to the transcoder 2 MAC.
				\item Send packets destined for the transcoder 2 MAC to transcoder 2.
				\item Send ARP requested for the transcoder IP to transcoder 2.
				\item Drop any packets received from transcoder 2.
			\end{itemize}
		\end{itemize}
	\item Wait a set time for the server to send out a ARP request to the transcoder, which will now be redirected to transcoder 2.
	\item Then in parallel:
		\begin{itemize}
			\item On switch 1:
			\begin{itemize}
				\item Disable transcoder 1 and the port it was connected to.
				\item Send packets from server to switch 2.
				\item Delete old stream duplication flows that are no longer required.
			\end{itemize}
			\item On switch 2:
			\begin{itemize}
				\item Send packets from transcoder 2 to the client.
				\item Delete old drop packet flows that and no longer required.
			\end{itemize}
		\end{itemize}
\end{enumerate}

It is important to note, that after the migration process has finished, there is no packet header manipulation being performed. This is essential to maintain the scalability of the system, as packet header manipulation places a considerable load on the switches, which might not be possible in real time with a large quantity of parallel streams. This is also means that the integrity of the layer 2 architecture is not diminished, since there are not two devices with the same MAC on the network even for a short amount of time. If this was not the case duplicate MAC addresses could cause issues in migration, where the machine is cloned along with the MAC address; this can cause confusion to networking devices, which are trying to identify locations of devices using source MAC learning.

\section{Results}
\label{sec:results}
The results presented here were collected over a span of 50 migrations for each migration type and were gathered using packet capture at the client. The packet capture was analysed to determine the packet delay between the last packet from the original transcoder (transcoder 1) and the first new packet from the new transcoder (transcoder 2). Since both transcoders utilise the same IP address, the source MAC address was used to differentiate between the two transcoders. There were some instances of overlap between the two transcoder streams where the packets had most likely been delayed at some point, but these overlaps were rare and only consisted of about 10-20 packets, which the playback system (VLC~\cite{vlc}) handled without issue.

Table \ref{table:migration_times} provides the mean migration times for the various types of migration recorded. It can clearly be determined from these results that OpenFlow aided migrations provide a significant improvement over standard stop and start migration techniques. Although the use of an ARP Flush on the server improves the performance of the non OpenFlow migration, it still does not show sufficient improvement to match that of the OpenFlow aided migrations. Note that in the case of OpenFlow, the ARP flush is not necessary and flushing the ARP table only adds to the delay while the server waits for the ARP.

It is important to note that the ARP Flushed results are shown only to provide an insight into how the ARP timeout on the server is one of largest contributing factors to the packet delay when using the non OpenFlow aided migrations. This delay is caused by the server not realising that the old transcoder is no longer active and a new transcoder with new MAC address has now taken over the transcoding at the same IP. However, it would not be feasible to use an ARP flush like this in a real world system as it would require access to the video server; whereas the system detailed here is being designed as a transparent system to the server and client, where apart from the transcoders physical migration, all the migration actions take place within the network with no modification to the client or server.

\begin{table}[t]
	\begin{center}
	  \begin{tabular}{ | c | c | c | c | c |}
	    \hline
	    \textbf{Migration Type} & \textbf{Mean Time} & \textbf{95\% CI} & \textbf{Min} & \textbf{Max} \\ \hline
	    125ms OF Aided & 0.0743 & $\pm$0.0184 & 0.00001 & 0.2482 \\ \hline
	    125ms Standard & 18.2414 & $\pm$2.0579 & 3.0056 & 37.7518 \\ \hline
	    250ms OF Aided & 0.0463 & $\pm$0.0153 & 0.00001 & 0.1308 \\ \hline
	    250ms Standard & 20.0841 & $\pm$5.3761 & 3.5516 & 42.9213 \\ \hline
	    ARP Flush OF Aided & 0.1231 & $\pm$0.0201 & 0.000002 & 0.2928 \\ \hline
	    ARP Flush Standard & 4.2093 & $\pm$0.1311 & 2.8771 & 7.1119 \\ \hline
	    OF Aided & 0.1058 & $\pm$0.0191 & 0.00001 & 0.2580 \\ \hline
	    Standard & 16.2127 & $\pm$2.3677 & 2.7545 & 37.3414 \\ \hline
	  \end{tabular}	  
	\end{center}
	\caption[Average Migration Times]{The mean migration times for each method, both with and without the aid of OpenFlow.}
	\label{table:migration_times}
\end{table}

\begin{figure}[t]
	\centering
	\adjincludegraphics[width=\columnwidth,keepaspectratio]{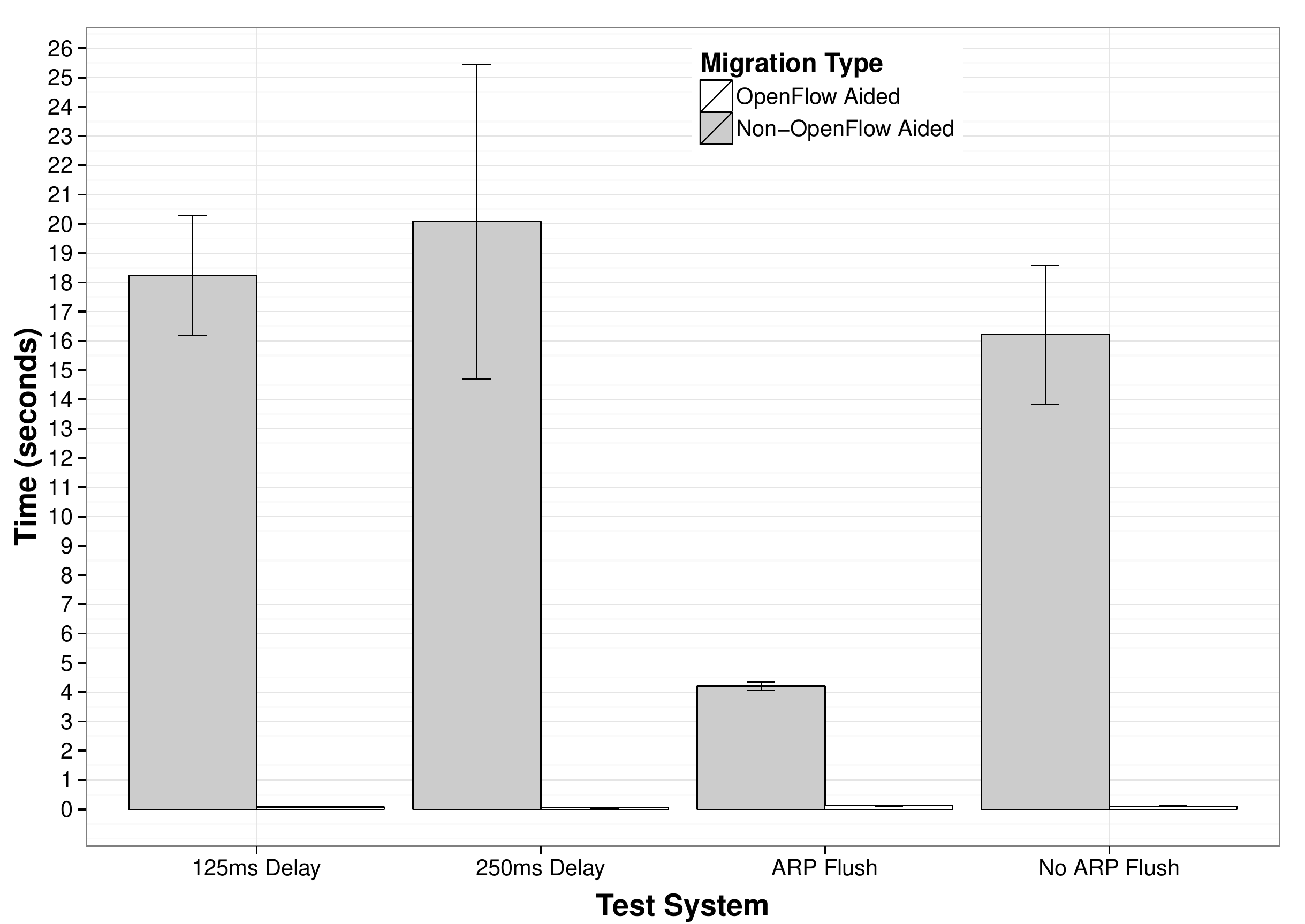}
	\caption{Delay between packets from new transcoder after old one is disconnected (95\% confidence intervals shown as error bars)}
	\label{fig:migration_delay}
\end{figure}

Fig. \ref{fig:migration_delay} shows the significant difference between the migration scenarios, with the OpenFlow aided migrations barely registering on the scale due to their sub second delay times. 

\begin{figure}[t]
	\centering
	\begin{tikzpicture}
	    \node[anchor=south west,inner sep=0] at (0,0) {\includegraphics[width=\columnwidth,keepaspectratio]{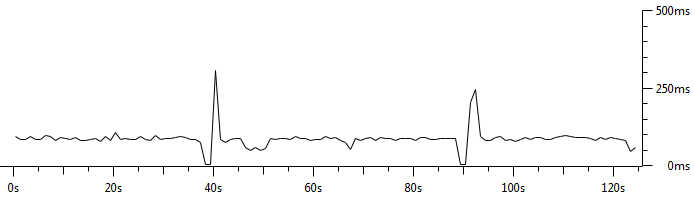}};
	    \node[anchor=north] at (1.3,3) {OpenFlow};
	    \draw[ultra thick,->] (1.3,2.5) -- (1.3,1);
	    \node[anchor=north] at (4.4,3) {OpenFlow};
	    \draw[ultra thick,->] (4.4,2.5) -- (4.4,1);
	    \node[anchor=north] at (2.6,3.8) {non-OpenFlow};
	    \draw[ultra thick,->] (2.6,3.3) -- (2.6,1.8);
	    \node[anchor=north] at (5.9,3.6) {non-OpenFlow};
	    \draw[ultra thick,->] (5.9,3.1) -- (5.9,1.6);
	\end{tikzpicture}
	\caption{Packet IO graph showing the time delta from the previous packet frame received at the client.}
	\label{fig:io_graph}
\end{figure}
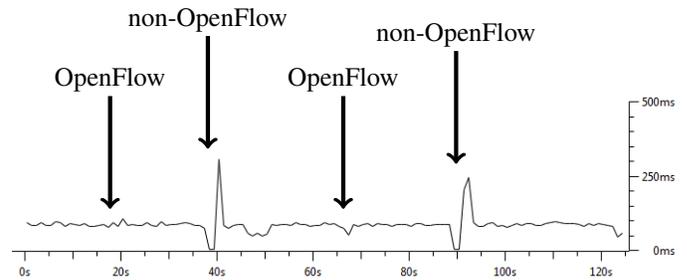

It can be seen clearly in Fig. \ref{fig:io_graph} that there is a significant improvement when OpenFlow is utilised to aid the migration process. There is minimal traffic disruption to the client during the OpenFlow migrations, in contrast with the standard migrations which produce a very large delay in the stream being re-established after the migration process. This delay in the packets can cause severe video playback issues, including freezing, lost frames and artefacts in the video; these issues would degrade the experience for the user which is something content providers aim to reduce.

\section{Conclusions and Future Work}
\label{sec:conclusions}
It is clear that the utilisation of OpenFlow to optimise transcoder migration is beneficial to the clients user experience. To migrate during streaming without it would most likely not be attempted, as it would introduce an unacceptable break in the transmission. It is also shown how the presented heuristic algorithm can provide a solution for the transcoder placement optimisation problem, while achieving a similar result to that of the presented GA but in a fraction of the time.

This provides the capability to optimise transcoder placement using the placement algorithm multiple times during a transmission, providing a highly optimised system throughout the length of the transmission, even with a client population shift. This principle can also be adapted for use with other scenarios, such as migrating transcoders based on reducing data center costs; this could involve moving transcoding resources around the globe to follow the day/night cycle or moving resources to cheaper under utilised data centers.

It has also been shown that the use of transcoders as application layer multicasters makes a significant contribution to reducing the network carried within a given network. Although this is not a new concept, it is important to note this fact alongside this system, since it is part of the optimisation strategy.

Although this paper provides a detailed overview of the research that has been undertaken, there are still areas which require more work. These include the integration of the designed control software into an OpenFlow controller application, allowing the system to function as a standard OpenFlow system without using extra tools. Another area which requires more insight is how the effects of more diverse demands affect the efficiency of the system, both during optimisation as well as during migration.

\appendices
\section{Control Software Design}
The control software used within this paper was developed as a threaded SSH application in Java, allowing concurrent control of devices such as the Open vSwitch instances running on the Pica8 P-3290 switches that were used. It should be noted that the system can function with any OpenFlow enabled switch if integrated within a controller application.

\def\arraystretch{2.2}
\begin{table*}[b]
	\begin{center}
	  \begin{tabular}{ | c | p{.65\paperwidth} | }
	    \hline
	    \textbf{Flow} & \textbf{FlowMod} \\ \hline
	    1 & \parbox{.65\paperwidth}{cookie=9998,in\_port=serverPORT,dl\_type=0x0800,nw\_src=serverIP,nw\_dst=transcoderIP \\ actions=output:sw1LinkPORT,mod\_dl\_dst:transcoder1MAC,output:transcoder1PORT} \\
	    \hline
	    2 & \parbox{.65\paperwidth}{cookie=9999,in\_port=serverPORT,dl\_type=0x0806,nw\_src=serverIP,nw\_dst=transcoderIP \\ actions=output:sw1LinkPORT} \\
	    \hline
	    3 & \parbox{.65\paperwidth}{cookie=9999,in\_port=sw2LinkPORT,dl\_type=0x0800,dl\_dst=transcoder1MAC,nw\_src=serverIP,nw\_dst=transcoderIP \\ actions=mod\_dl\_dst:transcoder2MAC,output:transcoder2PORT} \\ 
	    \hline
	    4 & \parbox{.65\paperwidth}{cookie=9999,in\_port=sw2LinkPORT,dl\_type=0x0800,dl\_dst=transcoder2MAC,nw\_src=serverIP,nw\_dst=transcoderIP \\ actions=output:transcoder2PORT} \\
	    \hline
	    5 & \parbox{.65\paperwidth}{cookie=9999,in\_port=sw2LinkPORT,dl\_type=0x0806,nw\_src=serverIP,nw\_dst=transcoderIP \\actions=output:transcoder2PORT} \\ 
	    \hline
	    6 & \parbox{.65\paperwidth}{cookie=9997,in\_port=transcoder2PORT,dl\_type=0x0800,nw\_src=transcoderIP,nw\_dst=clientIP \\ actions=} \\	    
	    \hline
	    7 & \parbox{.65\paperwidth}{cookie=9999,in\_port=transcoder2PORT,dl\_type=0x0800,nw\_src=transcoderIP,nw\_dst=clientIP \\ actions=output:clientPORT} \\
	    \hline
	    8 & \parbox{.65\paperwidth}{del-flows sw2Name cookie=9997/-1} \\
	    \hline
	    9 & \parbox{.65\paperwidth}{cookie=9999,in\_port=serverPORT,dl\_type=0x0800,nw\_src=serverIP,nw\_dst=transcoderIP \\ actions=output:sw1LinkPORT} \\
	    \hline
	    10 & \parbox{.65\paperwidth}{del-flows sw1Name cookie=9998/-1} \\
	    \hline
	  \end{tabular}	  
	\end{center}
	\caption[Flow Table]{Flow Mod details for the migration process.}
	\label{table:flow_table}
\end{table*}

\section*{Acknowledgement}

The authors would also like to thank the VISIONAIR project (project funded by the European Commission under grant agreement 262044) for providing UHD infrastructure for testing as a part of trans-national access and for the EPSRC supporting the first author under a Doctoral Training Account.

\ifCLASSOPTIONcaptionsoff
  \newpage
\fi

\bibliographystyle{IEEEtran}
\bibliography{bibliography}

\begin{IEEEbiography}[{\includegraphics[width=1in,height=1.25in,clip,keepaspectratio]{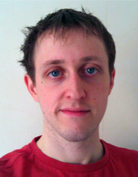}}]{Paul Farrow}
Paul Farrow graduated from the University of Essex in July 2011 and is currently set to complete his PhD by the end of 2015. His technical background is network optimisation, software engineering and software defined networking.
\end{IEEEbiography}%
\begin{IEEEbiographynophoto}{Martin Reed}
Martin Reed is a Senior Lecturer at the University of Essex. He received his PhD from University of Essex, UK, in 1998. After working as a research officer in the area of multimedia transmission over IP/ATM networks he was appointed as a lecturer at the University of Essex in 1998. His research interests include transmission of media, including ultra-high definition video, over networks, the control of transport networks, information centric networking and network security. He has been involved in a number of EPSRC, EU and Industrial projects in these areas and has held a Research Fellowship at BT. 
\end{IEEEbiographynophoto}%
\begin{IEEEbiographynophoto}{Maciej Glowiak}
Maciej Glowiak graduated from Poznan University of Technology in 2003 and then began working for the Poznan Supercomputing and Networking Center. His technical background entails network software engineering and parallel computing. Glowiak’s professional research interests include high-capacity networks, network monitoring, new protocols and future internet, new multimedia technology research, UHD video hardware and software development, among other areas. Glowiak is involved in a number of European network endeavors including GEANT and VISIONAIR. He is leading UHD-NET research and access activities in the VISIONAIR project. Since 2008 he has been involved into building UHD 3D nodes in Poland. Now, he works on new 8K 3D appliances.
\end{IEEEbiographynophoto}%
\begin{IEEEbiographynophoto}{Joe Mambretti}
Joe Mambretti is Director of the International Center for Advanced Internet Research at Northwestern University, which is developing digital communications for the 21st Century. The Center, which was created in partnership with a number of major high tech corporations, designs and implements large scale services and infrastructure for data intensive applications (metro, regional, national, and global). He is also Director of the Metropolitan Research and Education Network (MREN), an advanced high-performance network interlinking organization providing services in seven upper-Midwest states, and Director of the StarLight International/National Communications Exchange Facility in Chicago, which is based  on leading-edge optical technologies. With its research partners, iCAIR has established multiple major national and international network research testbeds, which are used to develop new architecture and technology for dynamically provisioned communication services and networks, including those based on lightpath switching.
\end{IEEEbiographynophoto}%
\end{document}